\documentclass[twocolumn,english,superscriptaddress,citeautoscript,showpacs,preprintnumbers,amsmath,amssymb,prb,floatfix,footinbib]{revtex4}
\usepackage[utf8x]{inputenc}
\usepackage[scaled=2]{helvet}
\usepackage[T1]{fontenc}
\usepackage[utf8]{luainputenc}
\usepackage{multirow}
\usepackage{textcomp}
\usepackage{amsmath}
\usepackage{amssymb}
\usepackage{graphicx}
\usepackage{subscript}
\usepackage{epstopdf}

\makeatletter

\providecommand{\tabularnewline}{\\}

\@ifundefined{textcolor}{}
{%
 \definecolor{BLACK}{gray}{0}
 \definecolor{WHITE}{gray}{1}
 \definecolor{RED}{rgb}{1,0,0}
 \definecolor{GREEN}{rgb}{0,1,0}
 \definecolor{BLUE}{rgb}{0,0,1}
 \definecolor{CYAN}{cmyk}{1,0,0,0}
 \definecolor{MAGENTA}{cmyk}{0,1,0,0}
 \definecolor{YELLOW}{cmyk}{0,0,1,0}
}

\usepackage[scaled=2]{helvet}\usepackage{amstext}

\makeatletter
\@ifundefined{textcolor}{}{\definecolor{BLACK}{gray}{0}\definecolor{WHITE}{gray}{1}\definecolor{RED}{rgb}{1,0,0}\definecolor{GREEN}{rgb}{0,1,0}\definecolor{BLUE}{rgb}{0,0,1}\definecolor{CYAN}{cmyk}{1,0,0,0}\definecolor{MAGENTA}{cmyk}{0,1,0,0}\definecolor{YELLOW}{cmyk}{0,0,1,0}}

\usepackage{dcolumn}
\usepackage{bm}
\usepackage{times}
\usepackage{hyperref}
\hypersetup{colorlinks=true,linkcolor=red,citecolor=blue}
\makeatother

\usepackage{babel}

\makeatother

\usepackage{babel}
\begin{document}

\title{Phase diagram of Eu magnetic ordering in Sn-flux-grown Eu(Fe$_{1-x}$Co$_{x}$)$_{2}$As$_{2}$ single crystals}

\author{W. T. Jin}

\email{w.jin@fz-juelich.de}

\affiliation{J\"{u}lich Centre for Neutron Science JCNS and Peter Gr\"{u}nberg Institut PGI, JARA-FIT, Forschungszentrum J\"{u}lich GmbH, D-52425 J\"{u}lich, Germany}

\affiliation{J\"{u}lich Centre for Neutron Science JCNS at Heinz Maier-Leibnitz Zentrum (MLZ), Forschungszentrum J\"{u}lich GmbH, Lichtenbergstraße 1, D-85747 Garching, Germany}

\author{Y. Xiao}

\affiliation{J\"{u}lich Centre for Neutron Science JCNS and Peter Gr\"{u}nberg Institut PGI, JARA-FIT, Forschungszentrum J\"{u}lich GmbH, D-52425 J\"{u}lich, Germany}

\author{Z. Bukowski }

\affiliation{Institute of Low Temperature and Structure Research, Polish Academy of Sciences, 50-422 Wroclaw, Poland}

\author{Y. Su}

\affiliation{J\"{u}lich Centre for Neutron Science JCNS at Heinz Maier-Leibnitz Zentrum (MLZ), Forschungszentrum J\"{u}lich GmbH, Lichtenbergstraße 1, D-85747 Garching, Germany}

\author{S. Nandi}

\affiliation{Department of Physics, Indian Institute of Technology, Kanpur 208016, India}

\author{A. P. Sazonov}

\affiliation{RWTH Aachen University, Institut f\"{u}r Kristallographie, D-52056 Aachen, Germany}

\affiliation{J\"{u}lich Centre for Neutron Science JCNS at Heinz Maier-Leibnitz Zentrum (MLZ), Forschungszentrum J\"{u}lich GmbH, Lichtenbergstraße 1, D-85747 Garching, Germany}

\author{M. Meven}

\affiliation{RWTH Aachen University, Institut f\"{u}r Kristallographie, D-52056 Aachen, Germany}

\affiliation{J\"{u}lich Centre for Neutron Science JCNS at Heinz Maier-Leibnitz Zentrum (MLZ), Forschungszentrum J\"{u}lich GmbH, Lichtenbergstraße 1, D-85747 Garching, Germany}

\author{O. Zaharko}

\affiliation{Laboratory for Neutron Scattering and Imaging, Paul Scherrer Institut, CH-5232 Villigen PSI, Switzerland}

\author{S. Demirdis}

\affiliation{J\"{u}lich Centre for Neutron Science JCNS at Heinz Maier-Leibnitz Zentrum (MLZ), Forschungszentrum J\"{u}lich GmbH, Lichtenbergstraße 1, D-85747 Garching, Germany}

\author{K. Nemkovski}

\affiliation{J\"{u}lich Centre for Neutron Science JCNS at Heinz Maier-Leibnitz Zentrum (MLZ), Forschungszentrum J\"{u}lich GmbH, Lichtenbergstraße 1, D-85747 Garching, Germany}

\author{K. Schmalzl}

\affiliation{J\"{u}lich Centre for Neutron Science JCNS at Institut Laue-Langevin (ILL), Forschungszentrum J\"{u}lich GmbH, Boite Postale 156, 38042 Grenoble Cedex 9, France}

\author{Lan Maria Tran}

\affiliation{Institute of Low Temperature and Structure Research, Polish Academy of Sciences, 50-422 Wroclaw, Poland}

\author{Z. Guguchia}

\affiliation{Laboratory for Muon Spin Spectroscopy, Paul Scherrer Institut, CH-5232 Villigen PSI, Switzerland}

\author{E. Feng}

\affiliation{J\"{u}lich Centre for Neutron Science JCNS at Heinz Maier-Leibnitz Zentrum (MLZ), Forschungszentrum J\"{u}lich GmbH, Lichtenbergstraße 1, D-85747 Garching, Germany}

\author{Z. Fu}

\affiliation{J\"{u}lich Centre for Neutron Science JCNS at Heinz Maier-Leibnitz Zentrum (MLZ), Forschungszentrum J\"{u}lich GmbH, Lichtenbergstraße 1, D-85747 Garching, Germany}

\author{Th. Br\"{u}ckel}

\affiliation{J\"{u}lich Centre for Neutron Science JCNS and Peter Gr\"{u}nberg Institut PGI, JARA-FIT, Forschungszentrum J\"{u}lich GmbH, D-52425 J\"{u}lich, Germany}

\affiliation{J\"{u}lich Centre for Neutron Science JCNS at Heinz Maier-Leibnitz Zentrum (MLZ), Forschungszentrum J\"{u}lich GmbH, Lichtenbergstraße 1, D-85747 Garching, Germany}

\begin{abstract}
The magnetic ground state of the Eu$^{2+}$ moments in a series of Eu(Fe$_{1-x}$Co$_{x}$)$_{2}$As$_{2}$ single crystals grown from the Sn flux has been investigated in detail by neutron diffraction measurements. Combined with the results from the macroscopic properties (resistivity, magnetic susceptibility and specific heat) measurements, a phase diagram describing how the Eu magnetic order evolves with Co doping in Eu(Fe$_{1-x}$Co$_{x}$)$_{2}$As$_{2}$ is established. The ground-state magnetic structure of the Eu$^{2+}$ spins is found to develop from the A-type antiferromagnetic (AFM) order in the parent compound, via the A-type canted AFM structure with some net ferromagnetic (FM) moment component along the crystallographic $\mathit{c}$ direction at intermediate Co doping levels, finally to the pure FM order at relatively high Co doping levels. The ordering temperature of Eu declines linearly at first, reaches the minimum value of 16.5(2) K around $\mathit{x}$ = 0.100(4), and then reverses upwards with further Co doping. The doping-induced modification of the indirect Ruderman-Kittel-Kasuya-Yosida (RKKY) interaction between the Eu$^{2+}$ moments, which is mediated by the conduction $\mathit{d}$ electrons on the (Fe,Co)As layers, as well as the change of the strength of the direct interaction between the Eu$^{2+}$ and Fe$^{2+}$ moments, might be responsible for the change of the magnetic ground state and the ordering temperature of the Eu sublattice. In addition, for Eu(Fe$_{1-x}$Co$_{x}$)$_{2}$As$_{2}$ single crystals with 0.10 $\leqslant$ $\mathit{x}$ $\leqslant$ 0.18, strong ferromagnetism from the Eu sublattice is well developed in the superconducting state, where a spontaneous vortex state is expected to account for the compromise between the two competing phenomena. 
\end{abstract}


\maketitle

\section{Introduction}

Iron-based superconductors discovered in 2008 \cite{Kamihara_08} have provided new opportunities to study the interplay between superconductivity (SC) and magnetism, as the SC in these new materials was found to emerge based on the suppression of static long-range ordered antiferromagnetism,\cite{Johnston_10, Dai_15} similar to that in unconventional cuprate superconductors.\cite{Tranquada_14} The ternary ``122'' $\mathit{A}$Fe$_{2}$As$_{2}$ (with $\mathit{A}$ = Ba, Sr, Ca, or Eu) parent compounds, which crystallize in the tetragonal ThCr$_{2}$Si$_{2}$ type structure, stand out as model systems for researches, since large, high-quality single crystals are available and various methods of tuning towards the SC have been reported.\cite{Johnston_10, Stewart_11}

EuFe$_{2}$As$_{2}$ is a special member of the ``122'' family, since the $\mathit{A}$ site is occupied by an $\mathit{S}$-state (orbital moment $\mathit{L}$ = 0) Eu$^{2+}$ rare earth ion. In a purely ionic picture, it has a 4$\mathit{f}$$^{7}$ electronic configuration
and a total electron spin $\mathit{S}$ = 7/2, corresponding to a theoretical effective magnetic moment of 7.94 $\mathit{\mu_{B}}$.\cite{Marchand_78} This compound exhibits a spin-density-wave (SDW) ordering of the itinerant Fe moments concomitant with a tetragonal-to-orthorhombic structural phase transition below 190 K. In addition, the localized Eu$^{2+}$ spins order below 19 K in an A-type antiferromagnetic (AFM) structure (ferromagnetic layers stacking antiferromagnetically along the $\mathit{c}$ direction).\cite{Herrero-Martin_09,Xiao_09} Superconductivity can be achieved in this system by suppressing the SDW ordering of Fe in the form of chemical substitution \cite{Jeevan_08,Ren_09,Jiang_09,Jiao_11,Jiao_13} or applying external pressure.\cite{Miclea_09,Terashima_09} For example,
by hole doping with partial substitution of K for Eu, Eu$_{0.5}$K$_{0.5}$Fe$_{2}$As$_{2}$ shows SC below the superconducting transition temperature$\mathit{T_{sc}}$ = 32 K.\cite{Jeevan_08} Isovalent substitution of P into the As sites can also give rise to the SC with $\mathit{T_{sc}}$ over 20 K.\cite{Ren_09,Jeevan_11} Nevertheless, for the electron-doped Eu(Fe$_{1-x}$Co$_{x}$)$_{2}$As$_{2}$ system, the reports about its physical properties remain quite controversial. Jiang $\mathit{et}$ $\mathit{al.}$ first discovered the onset of SC at $\mathit{T_{sc}}$ $\sim$ 21 K in the Eu(Fe$_{0.89}$Co$_{0.11}$)$_{2}$As$_{2}$ single crystals grown using (Fe, Co)As as the self-flux.\cite{Jiang_09} However, no zero-resistivity state was achieved in Ref. \onlinecite{Jiang_09} as well as in other self-flux-grown Eu(Fe$_{1-x}$Co$_{x}$)$_{2}$As$_{2}$ single crystals.\cite{Ying_10,Chen_10} To the best of our knowledge, so far the zero-resistivity state in Eu(Fe$_{1-x}$Co$_{x}$)$_{2}$As$_{2}$ was only realized in the single crystals grown using Sn as the flux, but with relatively lower superconducting transition temperature ($\mathit{T_{sc}}$ $\sim$ 10 K).\cite{Blachowski_11,Matusiak_11,Tran_12,Jin_13} In addition, the SDW order in the self-flux-grown crystals was found to be suppressed much faster upon Co doping than that in the Sn-flux-grown crystals.\cite{Ying_10,Matusiak_11} It is quite clear that the physical properties of the Eu(Fe$_{1-x}$Co$_{x}$)$_{2}$As$_{2}$ single crystals strongly depend on the growth techniques. Furthermore, in contrast to EuFe$_{2}$(As$_{1-x}$P$_{x}$)$_{2}$ system whose phase diagram was already thoroughly investigated,\cite{Guguchia_13, Jeevan_11, Zapf_11, Zapf_13, Xu_14} a specific phase diagram of Eu(Fe$_{1-x}$Co$_{x}$)$_{2}$As$_{2}$ describing how the magnetic order of the Eu$^{2+}$ moments develops with Co doping and how it is linked with the occurrence of SC has not been established yet.

Here we have performed comprehensive and systematic studies on single crystals of Eu(Fe$_{1-x}$Co$_{x}$)$_{2}$As$_{2}$ with different Co-doping levels grown from the Sn flux. We have microscopically investigated the evolution of the ground-state Eu$^{2+}$ magnetic order with Co doping by single-crystal neutron diffraction measurements, and have established the phase diagram of Sn-flux-grown Eu(Fe$_{1-x}$Co$_{x}$)$_{2}$As$_{2}$ single crystals based on both macroscopic and microscopic measurements.

\section{Experimental Details }

Single crystals of Eu(Fe$_{1-x}$Co$_{x}$)$_{2}$As$_{2}$ with $\mathit{x}$ = 0.014(2), 0.027(2), 0.053(2), 0.075(2), 0.100(4) and 0.180(5) were grown from the Sn-flux.\cite{Guguchia_11} The concentration of Co ($\mathit{x}$) was determined by wavelength dispersive spectroscopy (WDS) for the crystals with $\mathit{x}$ = 0.075(2) and 0.180(5). Considering that the compounds with $\mathit{x}$ = 0 (parent compound), $\mathit{x}$ = 0.075(2) and 0.180(5) undergo the structural transition at $\mathit{T_{s}}$ = 190 K (Ref. \onlinecite{Xiao_09}), 150 K (as presented below in Fig. 1) and 90 K (Ref.\onlinecite{Jin_13}), respectively, $\mathit{T_{s}}$ is believed to decrease linearly with the Co concentration in this system. The $\mathit{x}$ values for the other four crystals were determined using linear extrapolation method according to their $\mathit{T_{s}}$ values shown respectively in the resistivity measurements (Fig. 1). All crystals were platelike with dimensions up to 5 $\times$ 5 $\times$ 1 mm$^{3}$ with the $\mathit{c}$ axis perpendicular to their surfaces. The neutron diffraction measurements were performed on the hot-neutron four-circle diffractometer HEiDi \cite{Meven_15} and the diffuse scattering cold-neutron spectrometer DNS \cite{Su_15} at the Heinz Maier-Leibnitz Zentrum (Garching, Germany), the thermal-neutron four-circle diffractometer TriCS \cite{Schefer_00} at Paul Scherrer Institute (Villigen, Switzerland), and thermal-neutron two-axis diffractometer D23 at Institut Laue Langevin (Grenoble, France), respectively. The experimental conditions for different compositions are summarized in Table 1. 

\begin{table*}
\caption{The summary of the conditions for single-crystal neutron diffraction measurements on different compositions of Eu(Fe$_{1-x}$Co$_{x}$)$_{2}$As$_{2}$, and the comparison between the magnetic order temperature of Eu ($\mathit{T_{Eu}}$) determined from different experimental methods. The data for the parent compound ($\mathit{x}$ = 0) is based on available results from previous publications (Ref. \onlinecite{Xiao_09, Jiang_09_NJP, Jeevan_08}).}

\begin{ruledtabular} %
\begin{tabular}{c|c|ccccccc}
Experimental Method & $\mathit{x}$  & 0  & 0.014(2) & 0.027(2) & 0.053(2) & 0.075(2) & 0.100(4) & 0.180(5)\tabularnewline
\hline 
\multirow{4}{*}{Neutron diffraction } & crystal mass (mg) & 50 & 110 & 105 & 16 & 76 & 94 & 100\tabularnewline
 & Instrument & HEiDi & HEiDi and DNS & HEiDi & TriCS and DNS  & D23 & DNS & TriCS\tabularnewline
 & wavelength (Å) & 0.868 & 1.162 and 4.2 & 1.162 & 1.178 and 4.2 & 1.279 & 4.2 & 1.178\tabularnewline
 & $\mathit{T_{Eu}}$ (K) & 19.0(2) (Ref. \onlinecite{Xiao_09}) & 18.4(1) & 18.1(1) & 17.5(1) & 17.0(2) & 16.5(2) & 16.9(2)\tabularnewline
\hline 
\multicolumn{1}{c|}{Magnetization} & $\mathit{T_{Eu}}$ (K) & 19 (Ref. \onlinecite{Jiang_09_NJP}) & 18.5(1) & 18.1(1) & 17.6(1) & 17.0(1) & 16.5(1) & 17.0(1)\tabularnewline
\hline 
Specific heat & $\mathit{T_{Eu}}$ (K) & 19 (Ref. \onlinecite{Jeevan_08}) & 18.3(1) & 18.2(1) & 17.6(1) & 17.1(1) & 16.5(1) & 17.1(1)\tabularnewline
\end{tabular}\end{ruledtabular} 
\end{table*}

Although the absorption effect of Eu for the cold neutrons with $\lambda$ = 4.2 \AA at DNS is quite severe, the thin platelike shape of the chosen crystals together with the large moment size ($\sim$ 7 $\mathit{\mu_{B}}$) of the Eu$^{2+}$ spins made the neutron measurements feasible.\cite{Herrero-Martin_09,Jin_16} At DNS, the {[}0, 1, 0{]} direction of the crystal was aligned perpendicular to the horizontal scattering plane so that the (H, 0, L) reciprocal plane can be mapped out by rotating the sample. Throughout this paper, the orthorhombic notation (space group $Fmmm$) will be used for convenience. Crystals from the same batches were characterized by macroscopic measurements including the resistivity, heat capacity and magnetic susceptibility, using the Quantum Design physical property measurement system (PPMS) and Quantum Design magnetic property measurement system (MPMS).

\section{Experimental Results }

Figure 1 shows the temperature dependences of the normalized in-plane resistivity ($\rho_{ab}$) of Eu(Fe$_{1-x}$Co$_{x}$)$_{2}$As$_{2}$ single crystals with $\mathit{x}$ = 0.014(2), 0.027(2), 0.053(2), 0.075(2), 0.100(4), and 0.180(5), respectively. For the parent compound EuFe$_{2}$As$_{2}$ ($\mathit{x}$ = 0), an upturn around 190 K was observed in the $\rho(T)$ curve, due to the opening of a gap at the Fermi surface associated with the SDW order and structural transition.\cite{Jeevan_08} Upon Co doping, it is clear that the structural phase transition ($\mathit{T_{s}}$) gets gradually suppressed, as shown in Fig. 1. The signatures of superconductivity set in for $\mathit{x}$ = 0.075(2), as its resistivity descends abruptly below $\mathit{T_{SC}}$ = 11.4(7) K. For $\mathit{x}$ = 0.100(4), the
resistivity shows a drastic drop below $\mathit{T_{SC}}$ = 10.1(4) K (see the inset of Fig. 1) and almost reaches zero. For $\mathit{x}$ = 0.18, although the SDW order and structural transition are not completely suppressed (as confirmed by previous neutron diffraction measurement),\cite{Jin_13} the resistivity drops sharply below $\mathit{T_{SC}}$ = 8.3(3) K and finally a zero-resistance state is achieved. 

\begin{figure}
\centering{}\includegraphics{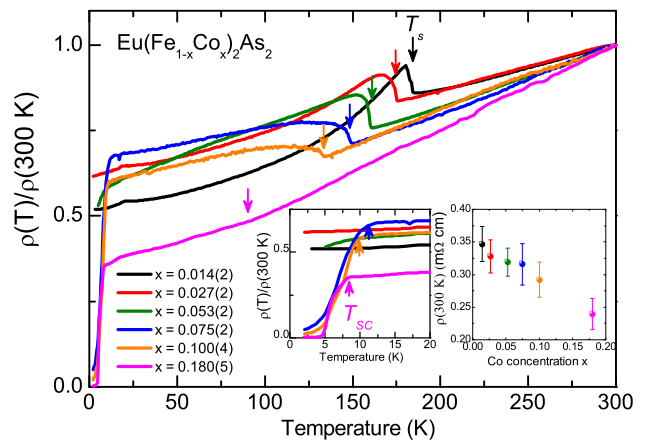}

\caption{Temperature dependences of the normalized in-plane resistivity ($\rho_{ab})$ of Eu(Fe$_{1-x}$Co$_{x}$)$_{2}$As$_{2}$ single crystals with $\mathit{x}$ = 0.014(2), 0.027(2), 0.053(2), 0.075(2), 0.100(4), and 0.180(5), respectively. The inset is an enlarged illustration of the resistivity below 20 K. $\mathit{T_{s}}$ (downward arrows) and $\mathit{T_{SC}}$ (upward arrows) mark the structural phase transition and the superconducting transition, respectively. The right inset shows the absolute resistivity values of different crystals at 300 K, in which the error bars are given by the uncertainties in the estimation of the geometric factors of the resistivity measurements.}
\end{figure}

Figure 2 summarizes the temperature dependences of the dc magnetic susceptibility ($\chi_{c}$) of the Eu(Fe$_{1-x}$Co$_{x}$)$_{2}$As$_{2}$ single crystals below 25 K, measured in the zero-field-cooling (ZFC) and field-cooling (FC) processes, respectively, in a small applied field (10 Oe or 30 Oe) parallel to the $\mathit{c}$-axis. For all six samples, the bifurcation between ZFC and FC curves develops below the ordering temperature of the Eu$^{2+}$ moments ($\mathit{T_{Eu}}$). With Co doping, $\mathit{T_{Eu}}$ decreases from 19 K in the parent compound ($\mathit{x}$ = 0), to 18.5(1) K for $\mathit{x}$ = 0.014(2), 18.1(1) K for $\mathit{x}$ = 0.027(2), 17.6(1) K for $\mathit{x}$ = 0.053(2), 17.0(1) K for $\mathit{x}$ = 0.075(2), finally to 16.5(1) K for $\mathit{x}$ = 0.100(4) (see also Table 1). Upon further Co doping, $\mathit{T_{Eu}}$ increases slightly to 17.0(1) K for $\mathit{x}$ = 0.180(5).\cite{Jin_13} Note that $\mathit{T_{Eu}}$ reaches 39 K for the end member EuCo$_{2}$As$_{2}$ with $\mathit{x}$ = 1, as reported by Ballinger $et\, al.$\cite{Ballinger_12} The tendency that $\mathit{T_{Eu}}$ initially gets suppressed with doping and then reverses upwards after reaching a minimum value around $\mathit{x}$ = 0.100(4) in Eu(Fe$_{1-x}$Co$_{x}$)$_{2}$As$_{2}$ is also confirmed by the heat capacity and neutron diffraction measurements, which will be presented below. It is noticeable that $\mathit{\chi_{c}}$ of the samples with $\mathit{x}$ = 0.100(4) and 0.180(5) exhibit a ``dip'' around 10 K and 8 K, respectively, as shown in Figs. 2(e) and 2(f), corresponding to the diamagnetic response due to the superconducting transition as suggested by the resistivity data shown in Fig.1. However, negative susceptibility is not achieved in these two samples due to the dominant effect of well-developed strong ferromagnetism of Eu at low temperature, as revealed by the neutron diffraction presented below and that reported in Ref. \onlinecite{Jin_13} . This is different from the ferromagnetic superconductors EuFe$_{2}$(As$_{0.85}$P$_{0.15}$)$_{2}$ and Eu(Fe$_{0.88}$Ir$_{0.12}$)$_{2}$As$_{2}$,\cite{Nandi_14,Jin_15} as for them the superconducting transition occurs at a temperature (25 K or 22 K) higher than the Curie temperature (19 K or 17 K) where it is easier to observe the negative susceptibility above 20 K in the paramagnetic background of Eu. For $\mathit{x}$ = 0.075(2), no such a dip in $\mathit{\chi_{c}}$ is discernible (Fig. 2(d)) although a sudden drop of its resistivity was also observed as shown in Fig. 1, indicating a small superconducting volume in this sample. Therefore, we only refer the crystals with $\mathit{x}$ = 0.100(4) and 0.180(5) as the superconducting samples.

\begin{figure}
\centering{}\includegraphics{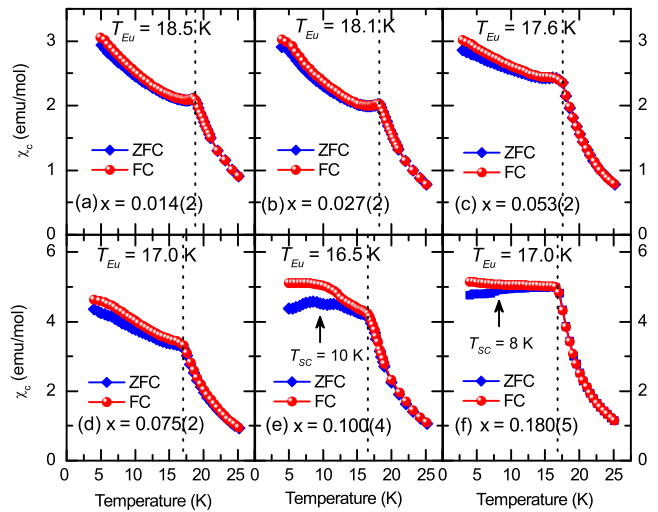}

\caption{Temperature dependences of the dc magnetic susceptibility ($\chi_{c})$ of Eu(Fe$_{1-x}$Co$_{x}$)$_{2}$As$_{2}$ single crystals below 25 K with $\mathit{x}$ = 0.014(2) (a), 0.027(2) (b), 0.053(2) (c), 0.075(2) (d), 0.100(4) (e), and 0.180(5) (f) (Ref. \onlinecite{Jin_13} ), measured in an applied field of 10 Oe ((a) to (e)) or 30 Oe (f) parallel to the $\mathit{c}$ direction in ZFC and FC processes, respectively. The dotted lines denote the ordering temperature of the Eu$^{2+}$ moments ($\mathit{T_{Eu}}$). The upward arrow in (e) and (f) marks the superconducting transition temperature ($\mathit{T_{SC}}$) of the samples with $\mathit{x}$ = 0.100(4) and 0.180(5), respectively.}
\end{figure}

The low-temperature molar specific heat of the Eu(Fe$_{1-x}$Co$_{x}$)$_{2}$As$_{2}$ single crystals, as shown in Fig. 3, clearly exhibits the evolution of the ordering temperature of the Eu$^{2+}$ moments ($\mathit{T_{Eu}}$) as a function of the Co doping level (see also Table 1). $\mathit{T_{Eu}}$, well determined from the peak position of the first derivative of the specific heat to temperature (-$\mathit{dC/dT}$, the inset of Fig. 3), gets suppressed to the minimum value of 16.5(1) K for $\mathit{x}$ = 0.100(4) and then increases again to 17.0(1) K for $\mathit{x}$ = 0.180(5), consistent with the results from the magnetic susceptibility measurements presented above. However, due to dominant contributions from both the phonons and the magnetic order of the Eu$^{2+}$ moments to the specific heat at low temperature, the expected anomaly at the superconducting transition for $\mathit{x}$ = 0.100(4) and 0.180(5) can be hardly resolved.

\begin{figure}
\centering{}\includegraphics{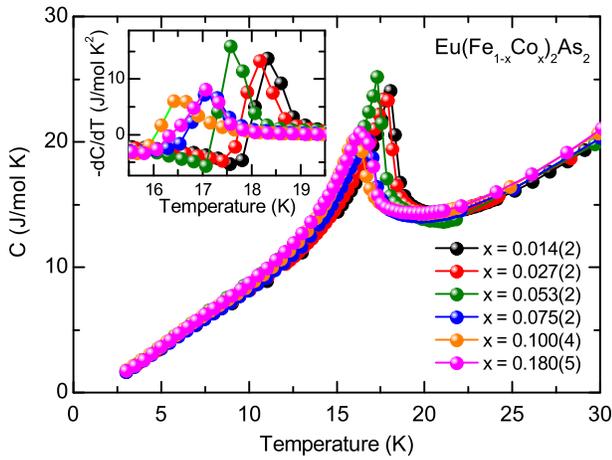}

\caption{The low-temperature molar specific heat of Eu(Fe$_{1-x}$Co$_{x}$)$_{2}$As$_{2}$ single crystals as a function of the Co doping level. The inset shows the first derivative of the specific heat to the temperature (-$\mathit{dC/dT}$) around the ordering temperature of the Eu$^{2+}$ moments ($\mathit{T_{Eu}}$).}
\end{figure}

The ground state magnetic structures of the Eu$^{2+}$ spins in Eu(Fe$_{1-x}$Co$_{x}$)$_{2}$As$_{2}$ with different Co doping levels are comprehensively investigated by single-crystal neutron diffraction and illustrated in Fig. 4, which will be discussed in details below.

\begin{figure}
\centering{}\includegraphics[width=1\columnwidth]{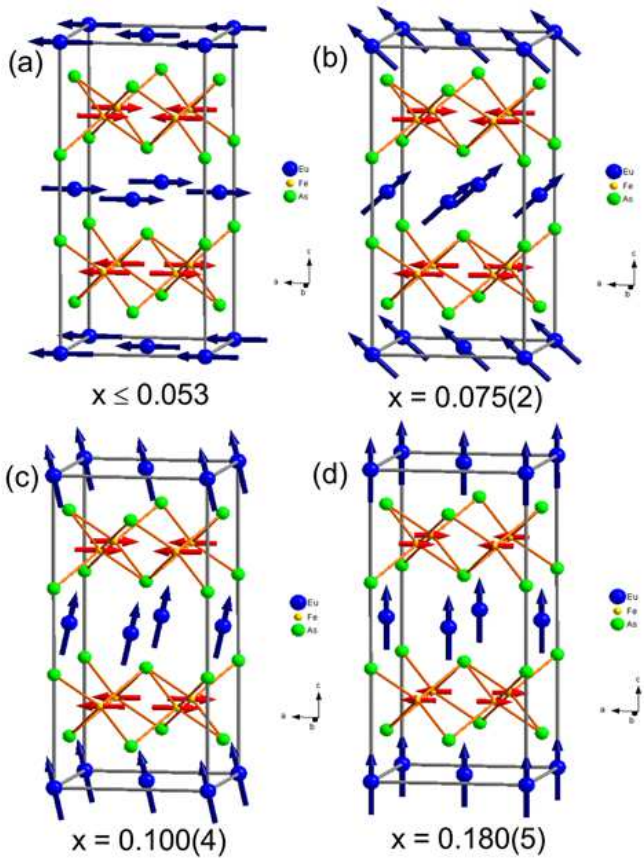}

\caption{The ground-state magnetic structure of Eu(Fe$_{1-x}$Co$_{x}$)$_{2}$As$_{2}$ with $\mathit{x}$$\leqslant$ 0.053 (a), $\mathit{x}$ = 0.075(2) (b), $\mathit{x}$ = 0.100(4) (c) and $\mathit{x}$ = 0.180(5) (d) Ref. \onlinecite{Jin_13} , respectively, as determined by single-crystal neutron diffraction measurements. The Eu$^{2+}$ spins in (b) and (c) are canted out of the $\mathit{ab}$ plane with an angle of 23.8(6)$^{\circ}$ (at 2 K) and \textasciitilde{}65 $^{\circ}$ (at 4 K), respectively.}
\end{figure}

As shown in Fig. 5, for the first three compositions with the Co doping level ($\mathit{x}$) less than 6\%, the ground-state magnetic structure of the Eu$^{2+}$ moments resembles that of the parent compound. At the base temperature, the localized Eu$^{2+}$ spins order in the A-type AFM structure (ferromagnetic $\mathit{ab}$ planes stacking antiferromagnetically along the $\mathit{c}$ direction, see Fig. 4(a)) and there is no net ferromagnetic component along the $\mathit{c}$ axis, since there is no magnetic contribution superimposing on the
(2, 0, 0) nuclear reflection, as shown in Fig. 5(b), 5(e) and 5(h), which is most sensitive to the possible ferromagnetic ordering of the Eu$^{2+}$ moments. In order to determine the antiferromagnetic transition temperature ($\mathit{T_{N}}$), rocking-curve scans of the (0, 0, 3) magnetic reflection were performed at different temperatures, as shown in Fig. 5(a), 5(d) and 5(g). The temperature dependences of the integrated intensity (at HEIDI and TriCS) or peak intensity (at DNS) of the (0, 0, 3) reflection are plotted in Fig. 5(c), 5(f) and 5(i), for $\mathit{x}$ = 0.014(2), 0.027(2) and 0.053(2), respectively. It is evident that $\mathit{T_{N}}$ (marked by the dashed lines) gets suppressed continuously upon the Co doping, from 18.4(1) K for $\mathit{x}$ = 0.014(2), to 18.1(1) K for $\mathit{x}$ = 0.027(2), then to 17.5(1) K for $\mathit{x}$ = 0.053(2), well consistent with that determined from both magnetic susceptibility and specific heat measurements. 

\begin{figure*}
\centering{}\includegraphics[width=1\textwidth]{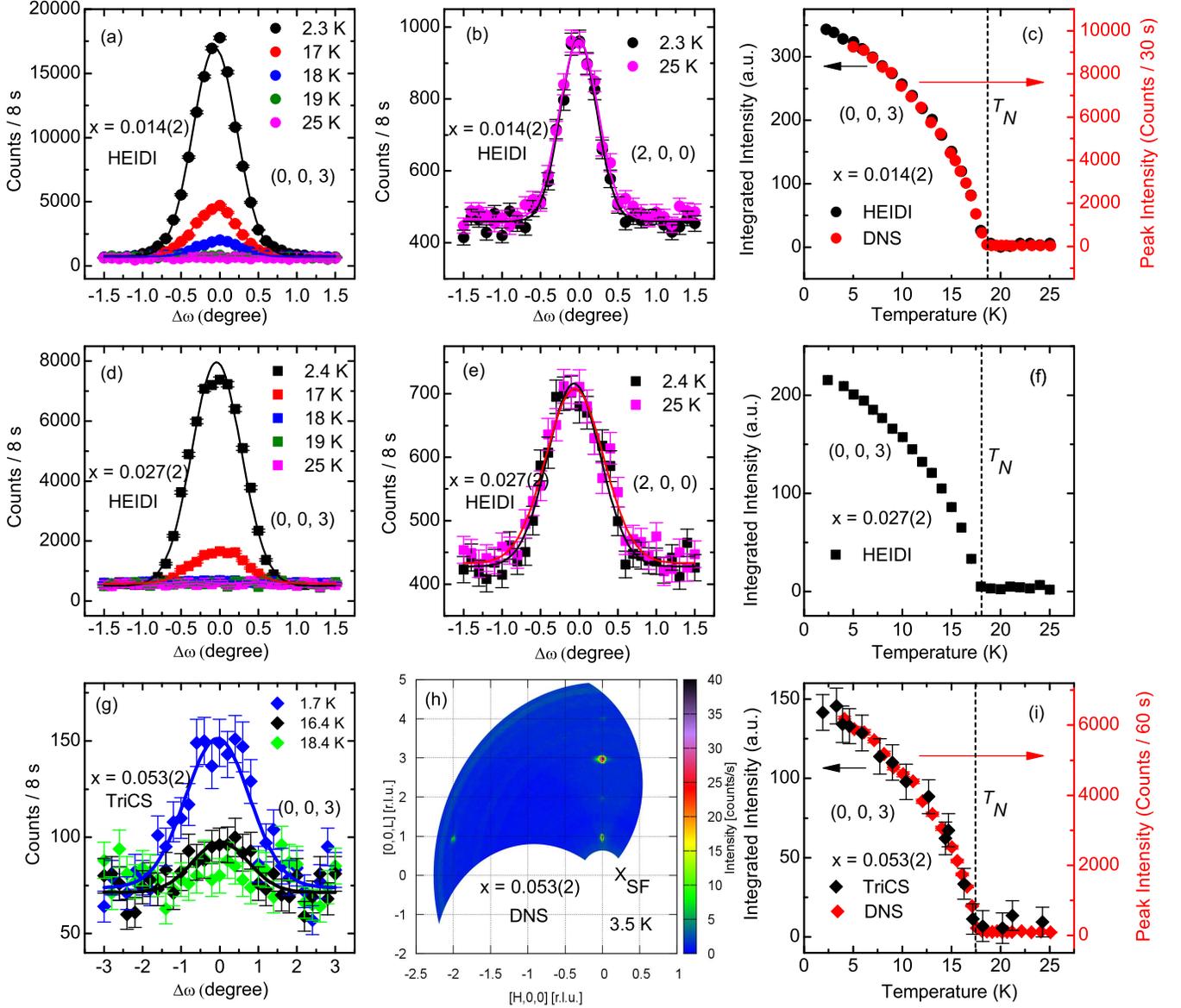}

\caption{The summary of neutron diffraction measurements on Eu(Fe$_{1-x}$Co$_{x}$)$_{2}$As$_{2}$ single crystals with $\mathit{x}$ = 0.014(2) (a-c), 0.027(2) (d-f), 0.053(2) (g-h), respectively, suggesting the A-type AFM structure as the magnetic ground state of the Eu$^{2+}$ spins. {[}(a), (d), (g){]} The rocking-curve scans of the (0, 0, 3) magnetic reflection performed on diffractometer HEiDi (a, d) or TriCS (g) at different temperatures for $\mathit{x}$ = 0.014(2) (a), $\mathit{x}$ = 0.027(2) (d), and $\mathit{x}$ = 0.053(2) (g), respectively. The solid curves represent the fits using the Gaussian profiles. {[}(b), (e){]} The rocking-curve scans of the (2, 0, 0) nuclear reflection performed on diffractometer HEiDi at the base temperature and 25 K, for $\mathit{x}$ = 0.014(2) (b) and $\mathit{x}$ = 0.027(2) (e), respectively. (h) The contour map in the (H, 0, L) reciprocal plane for $\mathit{x}$ = 0.053(2) at T = 3.5 K obtained via polarized neutron diffraction at spectrometer DNS with the neutron polarization parallel to the scattering vector $\mathit{Q}$ ($\mathit{x}$ polarization). For $\mathit{x}$ polarization, the intensity in the spin-flip (SF) channel solely arises from the magnetic reflections.\cite{Scharpf_93} {[}(c), (f), (i){]} The temperature dependences of the integrated intensity (at HEiDi and TriCS) or peak intensity (at DNS) of the (0, 0, 3) magnetic reflection. The vertical dashed lines mark the antiferromagnetic ordering temperature ($\mathit{T_{N}}$) of the Eu$^{2+}$ moments for each composition.}
\end{figure*}

With further increase of the Co doping level, the magnetic ground state of the Eu$^{2+}$ spins starts to differ from the A-type AFM structure. For the single crystal with $\mathit{x}$ = 0.075(2), as shown in Fig. 6(b), rocking-curve scans on diffractometer D23 reveal that the (2, 0, 0) reflection gets remarkably enhanced at 2 K, indicating considerable ferromagnetic (FM) contribution from the Eu$^{2+}$ moments. However, they are not purely ferromagnetically aligned along a unique crystallographic direction, as was observed for heavily doped crystal with $\mathit{x}$ = 0.180(5),\cite{Jin_13} since the (0, 0, 3) antiferromagnetic reflection is still present at 2 K (Fig. 6(a)). The integrated intensities of (0, 0, 3) and (2, 0, 0) peaks are plotted as a function of temperature in Fig. 6(c). Upon cooling, the antiferromagnetic and ferromagnetic contribution set in below the same temperature ($\mathit{T_{N}}$ = $\mathit{T_{C}}$ = 17.0(2) K), which is slightly lower compared to $\mathit{T_{N}}$ = 17.5(1) K for $\mathit{x}$ = 0.053(2). The integrated intensities of a set of nuclear and strong magnetic reflections collected at 25 K and 2 K, respectively, were refined using FULLPROF\cite{Rodriguez-Carvajal_93} after necessary absorption correction using DATAP.\cite{Coppens_65} With an A-type canted AFM model, in which the Eu$^{2+}$ moments on adjacent Eu layers order antiferromagnetically along the orthorhombic $\mathit{a}$ axis but yield a net ferromagnetic component along the $\mathit{c}$ axis (see Fig. 4(b)), the intensities of the collected magnetic reflections can be well fitted. The Eu$^{2+}$ spins are found to be canted with an angle of 23.8(6)$^{\circ}$ out of the $\mathit{ab}$ plane with the moment size of 6.22(3) $\mathit{\mu_{B}}$. Using the same Eu$^{2+}$ moment size, we calculated the intensities of the magnetic reflections for the A-type AFM structure or the pure FM structure along the $\mathit{c}$ axis. As shown in Fig. 6(e), both models can not account for all the observed magnetic reflections for $\mathit{x}$ = 0.075(2) while the A-type canted AFM structure shown in Fig. 4(b) can reproduce the observed intensities of all reflections. Moreover, in Fig. 6(c), the integrated intensities of the (2, 0, 0) and (0, 0, 3) reflections show different temperature dependences, suggesting that the canting angle of the Eu$^{2+}$ moments might change with temperature. After subtracting the nuclear contributions, we have calculated the ratio of the net magnetic scattering intensities between the (2, 0, 0) and (0, 0, 3) reflections, $\mathit{R=I_{M}}(2,0,0)/I_{M}(0,0,3)$, and the canting angle of the Eu$^{2+}$ moments, respectively. As shown in Fig. 6(d), below the ordering temperature of the Eu$^{2+}$ moments, the $\mathit{R}$ value increases upon cooling, corresponding to a smooth rotation of the Eu$^{2+}$ spins out of the $\mathit{ab}$ plane with temperature, from \textasciitilde{}20$^{\circ}$ around 15 K to \textasciitilde{}25$^{\circ}$ at the base tempearture. 

\begin{figure}
\centering{}\includegraphics{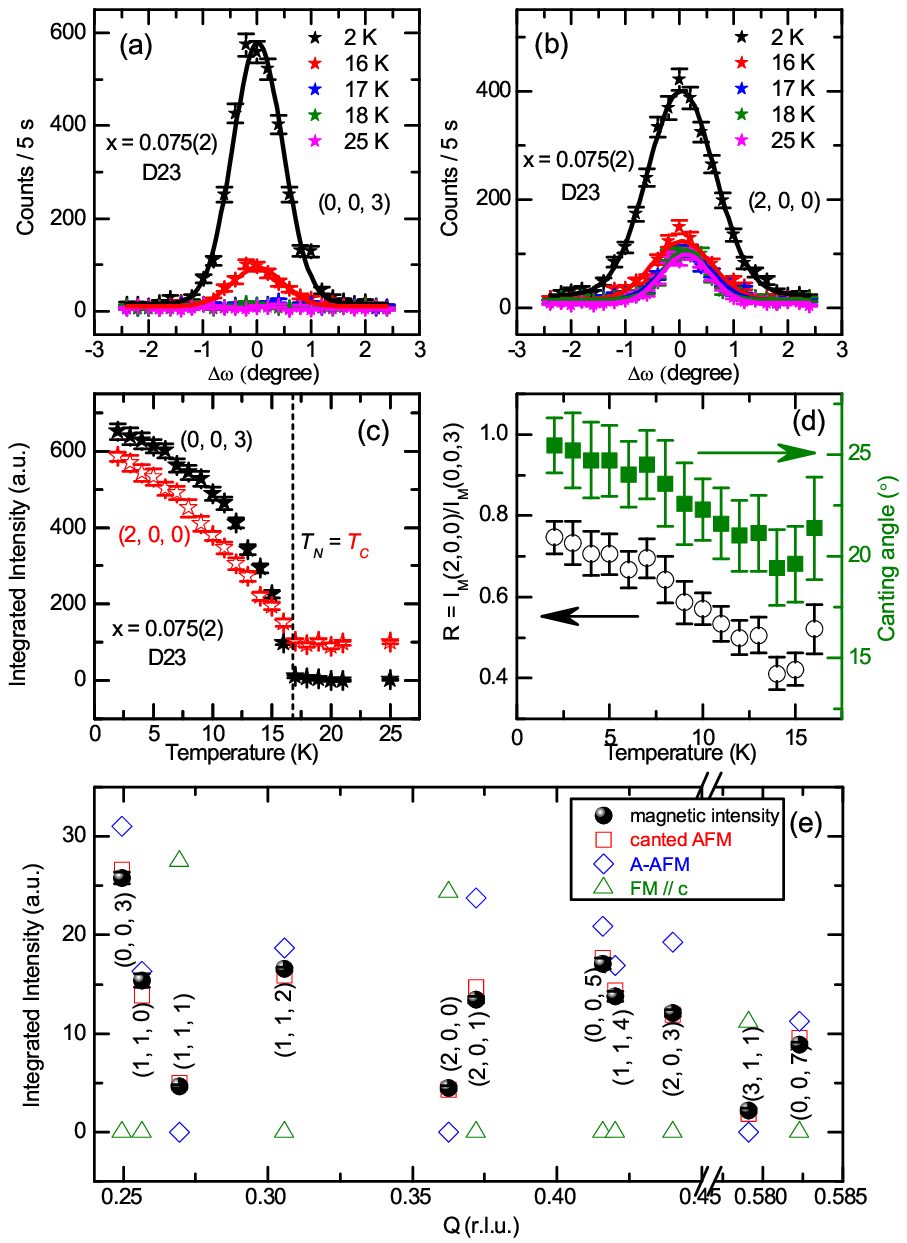}

\caption{{[}(a), (b){]} The rocking-curve scans of the (0, 0, 3) and (2, 0, 0) reflections, respectively, performed at diffractometer D23 at different temperatures for the Eu(Fe$_{1-x}$Co$_{x}$)$_{2}$As$_{2}$ single crystal with $\mathit{x}$ = 0.075(2). The solid curves represent the fits using the Gaussian profiles. (c) The temperature dependences of the integrated intensity of the (0, 0, 3) and (2, 0, 0) reflection, respectively, for $\mathit{x}$ = 0.075(2). The vertical dashed lines mark the same onset temperature for the antiferromagnetic and ferromagnetic contribution ($\mathit{T_{N}}$ = $\mathit{T_{C}}$ = 17.0(2) K). (d) The temperature dependences of the ratio of the net magnetic scattering intensities between the (2, 0, 0) and (0, 0, 3) reflections, $\mathit{R=I_{M}}(2,0,0)/I_{M}(0,0,3)$, and the canting angle of the Eu$^{2+}$ moments, respectively.(e) Comparison between the observed intensities of the collected magnetic reflections at 2 K (black spheres), the fitted intensities using the A-type canted AFM structure (red squares), the calculated intensities using the A-type AFM structure (blue diamonds) or the pure FM structure along the $\mathit{c}$ axis (green triangles), respectively. The observed intensities for (1, 1, 1), (2, 0, 0) and (3, 1, 1) reflections are the net magnetic contributions at 2 K after subtracting the nuclear contributions at 25 K. }
\end{figure}

Futhermore, the magnetic ground state of superconducting Eu(Fe$_{1-x}$Co$_{x}$)$_{2}$As$_{2}$ with $\mathit{x}$ = 0.100(4) ($\mathit{T_{SC}}$ = 10 K) was investigated by polarized neutron diffraction on spectrometer DNS. The polarization of incident neutrons was aligned along the {[}0, 1, 0{]} direction of the crystal, perpendicular to the horizontal scattering plane ($\mathit{z}$-polarization). With this configuration, the scattering cross-sections for the spin-flip (SF) and non-spin-flip (NSF) processes read as\cite{Scharpf_93}
\begin{equation}
\left(\frac{d\sigma}{d\Omega}\right)_{Z}^{SF}\propto M_{\bot Y}^{*}M_{\bot Y}+\frac{2}{3}I_{SI},
\end{equation}

and
\begin{equation}
\left(\frac{d\sigma}{d\Omega}\right)_{Z}^{NSF}\propto M_{\bot Z}^{*}M_{\bot Z}+N^{*}N+\frac{1}{3}I_{SI},
\end{equation}
respectively, where $M_{\bot Y}^{*}M_{\bot Y}$ and $M_{\bot Z}^{*}M_{\bot Z}$ are the components of the magnetization parallel and perpendicular to the scattering plane, respectively, with $\mathit{Y}$ being perpendicular to the scattering wavevector $\mathit{Q}$. $N^{*}N$ denotes the coherent nuclear scattering, and $I_{SI}$ denotes the total spin incoherent scattering. 

Figures 7(a) shows a reciprocal space contour map at T = 3.5 K obtained for the single crystal of Eu(Fe$_{1-x}$Co$_{x}$)$_{2}$As$_{2}$ with $\mathit{x}$ = 0.100(4) in the SF channel under $\mathit{z}$-polarization. The (0, 0, 1), (0, 0, 3) and (0, 0, 5) reflections arise from the antiferromagnetic alignment of the Eu$^{2+}$ moments on adjacent Eu layers along the crystallographic $\mathit{a}$ axis, similar to those observed for other samples with smaller Co concentrations as described above. However, the strong nuclear Bragg reflections (0,
0, 2) and (0, 0, 4), which should only appear in the NSF channel, become also visible in the SF channel. This results from the depolarization of the neutrons due to strong ferromagnetism from the Eu as explained below and thereby the imperfect flipping ratio correction. Fig. 7(b)
presents the integrated intensities of the (0, 0, 2) reflection in the SF (orange circles) and NSF channels (green circles) as a function of temperature. The intensity in the SF channel exhibits a order-parameter-like behavior below $\sim$16.5 K and that in the NSF channel drops continuously below the same temperature upon cooling, clearly revealing the depolarization of neutrons below 16.5 K. The total intensity of (0, 0, 2) (blue circles), which is the sum of the intensities in both channels, however, stays constant with temperature within the error bars, excluding possible ferromagnetic contribution on (0, 0, 2) reflection. In addition, both the (2, 0, 0) and (2, 0, 2) reflections appear in the SF channel, although they seem to be weaker compared with the (0, 0, L) reflections due to the stronger absorption effect. Since the nuclear scattering for (2, 0, 0) is extremely weak, the appearance of it in the SF channel can not be attributed to the neutron-depolarization effect but should be intrinsic. This means that there is some net ferromagnetic component of the Eu$^{2+}$ moments along the $\mathit{c}$-axis, which depolarizes the neutrons. Therefore, the magnetic ground state of Eu(Fe$_{1-x}$Co$_{x}$)$_{2}$As$_{2}$ with $\mathit{x}$ = 0.100(4) is also a canted AFM structure, resembling that of $\mathit{x}$ = 0.075(2). Upon cooling, the antiferromagnetic
and ferromagnetic contribution again set in below a same temperature ($\mathit{T_{N}}$ = $\mathit{T_{C}}$ = 16.5(2) K), as determined from the temperature dependences of the (0, 0, 3) and (2, 0, 0) reflections shown in Fig. 7(c). In contrast to the compound with $\mathit{x}$ = 0.075(2), the integrated intensities of the two reflections show very similar temperature dependences. The ratio of the magnetic scattering intensities in the SF channel between the (2, 0, 0) and (0, 0, 3) reflections, $\mathit{R=I_{ZSF}}(2,0,0)/I_{ZSF}(0,0,3)$, was calculated at different temperatures below the transition and shown in Fig. 7(d). The $\mathit{R}$ value stays almost constant within the error bars, indicating a stable canting angle of the Eu$^{2+}$ moments with the change of the temperature. The canting angle of the Eu$^{2+}$ spins out of the $\mathit{ab}$ plane can be estimated according to the magnitude of the net ferromagnetic contribution on the nuclear scattering part of the (2, 0, 2) reflection. The rocking-curve scans of the (2, 0, 2) reflection at 4.1 K and 25 K in the SF and NSF channels are shown in Fig. 7(e) and 7(f), respectively. It is evident that the (2, 0, 2) reflection, which has a strong contribution from the nuclear scattering, also suffers from the depolarization effect. By suming the integrated intensity in the SF and NSF channels, the total integrated intensity of the (2, 0, 2) reflection is plotted as a function of the temperature in Fig. 7(g). In contrast to the (0, 0, 2) reflection {[}Fig. 7(b){]}, the total intensity of (2, 0, 2) increases by$\sim$ 17.5\% below 16.5 K, due to ferromagnetic ordering of the Eu$^{2+}$
moments and displays a typical behavior expected for a magnetic order parameter. Assuming that the moment size of the Eu$^{2+}$ spins is 6.2 $\mathit{\mu_{B}}$, same as the value for $\mathit{x}$ = 0.075(2) (see the text above) and $\mathit{x}$ = 0.180(5) (see Ref. \onlinecite{Jin_13} ), the ratio between the ferromagnetic contribution and the nuclear scattering for the (2, 0, 2) reflection within the A-type canted AFM model can be quantitatively calculated as a function of the canting angle. As shown in Fig. 7(h), a ratio of 17.5\% (black sphere) corresponds to a canting angle of $\sim$65$^{\circ}$ off the $\mathit{ab}$ plane. The ground state magnetic structure of Eu(Fe$_{1-x}$Co$_{x}$)$_{2}$As$_{2}$ with $\mathit{x}$ = 0.100(4) determined here is illustrated in Fig. 4(c).

\begin{figure}
\centering{}\includegraphics{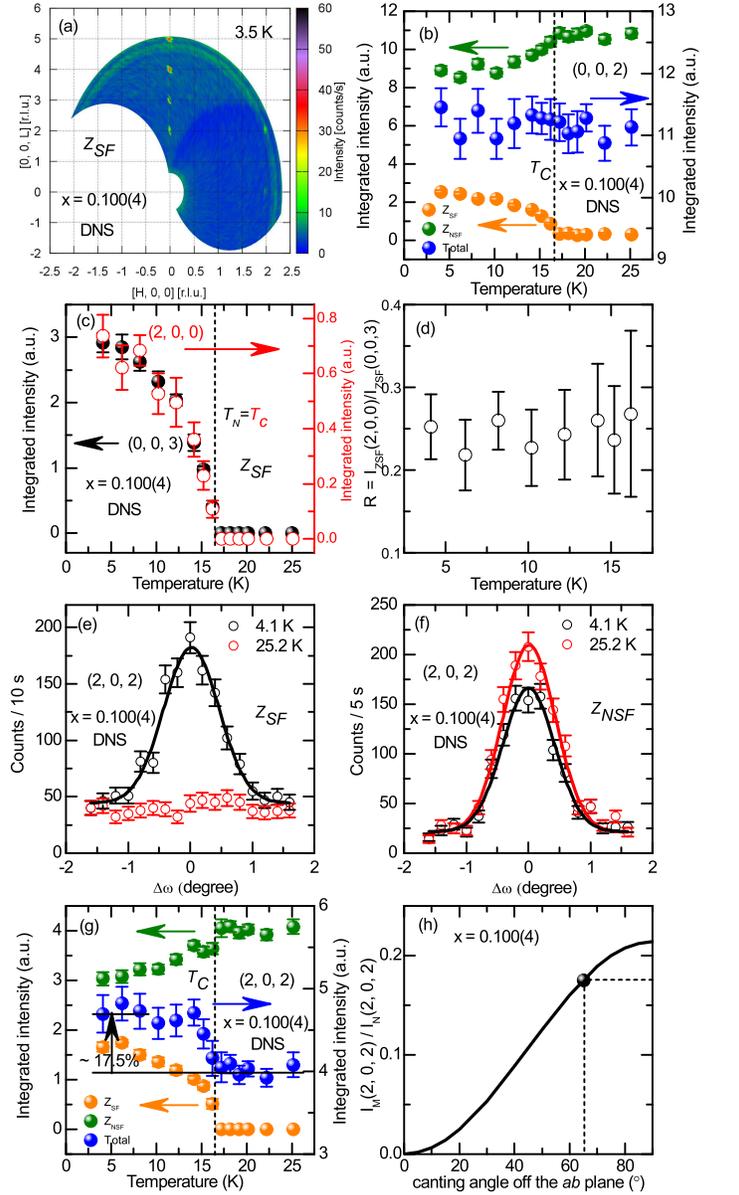}

\caption{(a) The contour map in the (H, 0, L) reciprocal plane for Eu(Fe$_{1-x}$Co$_{x}$)$_{2}$As$_{2}$ ($\mathit{x}$ = 0.100(4)) obtained at T = 3.5 K via polarized neutron diffraction on spectrometer DNS with $\mathit{z}$ polarization. (b) The temperature dependences of the SF-channel (orange circles), NSF-channel (green), and the total (blue) integrated intensities of the (0, 0, 2) reflection. (c) The temperature dependence of the integrated intensity of the (0, 0, 3) and (2, 0, 0) reflections, respectively. The vertical dashed lines mark the same onset temperature for the antiferromagnetic and ferromagnetic contribution ($\mathit{T_{N}}$ = $\mathit{T_{C}}$ = 16.5(2) K). (d) The temperature dependence of the ratio of the magnetic scattering intensities in the SF channel between the (2, 0, 0) and (0, 0, 3) reflections, $\mathit{R=I_{ZSF}}(2,0,0)/I_{ZSF}(0,0,3)$. {[}(e), (f){]} The rocking-curve scans of the (2, 0, 2) reflection at 4.1 K and 25 K in the SF and NSF channel, respectively. The solid curves represent the fits using the Gaussian profiles. (g) The temperature dependences of the SF-channel (orange circles), NSF-channel (green), and the total (blue) integrated intensities of the (2, 0, 2) reflection.(h) The ratio between the ferromagnetic contribution and the nuclear scattering for the (2, 0, 2) reflection within the A-type canted AFM model (solid line) calculated as a function of the canting angle, from which the canting angle for $\mathit{x}$ = 0.100(4) is determined to be $\sim$65$^{\circ}$ off the $\mathit{ab}$ plane.}
\end{figure}

The ground-state magnetic structure of superconducting Eu(Fe$_{1-x}$Co$_{x}$)$_{2}$As$_{2}$ with $\mathit{x}$ = 0.180(5) ($\mathit{T_{SC}}$ = 8 K) was determined in Ref. \onlinecite{Jin_13}  by unpolarized single-crystal diffraction measurements on diffractomter TriCS. The antiferromagnetic order of the Eu$^{2+}$ spins observed in other samples with smaller Co concentration as presented above was found to be completely suppressed for $\mathit{x}$ = 0.180(5) and a pure ferromagnetic alignment of the Eu$^{2+}$ moments along the $\mathit{c}$ direction was revealed there well below the Curie temperature $\mathit{T_{C}}$ = 17 K, as shown in Fig. 4(d). 

In Summary, Figure 4 illustrates how the ground-state magnetic structure of Sn-flux grown Eu(Fe$_{1-x}$Co$_{x}$)$_{2}$As$_{2}$ single crystals evolves with the increase of the Co concentration. For $\mathit{x}$$\leqslant$ 0.053, the Eu$^{2+}$ moments keep the alignment within the $\mathit{ab}$ plane in the A-type AFM order. Upon further increase of the Co doping level, the Eu$^{2+}$ spins start to rotate toward the $\mathit{c}$ axis within the $\mathit{ac}$ plane, exhibiting the A-type canted AFM structure with a net FM moment component along the $\mathit{c}$ direction. At base temperature, the canting angle of the Eu$^{2+}$ spins out of the $\mathit{ab}$ plane increases with Co doping, from 23.8(6)$^{\circ}$ for $\mathit{x}$ = 0.075(2), to $\sim$65$^{\circ}$ for $\mathit{x}$ = 0.100(4), finally to 90$^{\circ}$ for $\mathit{x}$ = 0.180(5) showing a pure FM order. This development of the magnetic ground state of localized Eu$^{2+}$ moments with Co doping, starting from the A-type AFM order, through the A-type canted AFM structure, to a pure FM order, is quite similar to the behavior observed in the perovskite magnanese oxides La$_{1-x}$Sr$_{x}$MnO$_{3}$ two decades ago.\cite{Kawano_96} The A-type canted AFM structure with some net FM magnetic component was reported there to be the magnetic ground state for the compositions with intermediate Sr doping level. However, the AFM and FM components display well seperated ordering temperatures in La$_{1-x}$Sr$_{x}$MnO$_{3}$, in contrast to the coincident AFM and FM transitions observed in Eu(Fe$_{1-x}$Co$_{x}$)$_{2}$As$_{2}$ with $\mathit{x}$ = 0.075(2) and $\mathit{x}$ = 0.100(4). 

In addition, similar to other doped 122 compounds, the SDW transition of the Fe moments is gradually suppressed by Co doping, as shown in Fig. 8. The peak intensity of the (1, 0, 3) or (1, 2, 1) magnetic reflection, measured on DNS or TriCS, is used as the order parameter associated with the Fe-SDW transition for $\mathit{x}$ = 0.014(2) and 0.180(5)\cite{Jin_13}, respectively. For $\mathit{x}$ = 0.027(2) and 0.075(2), the integrated intensity of the (1, 0, 3) reflection measured on HEiDi or D23, respectively, is used instead.\cite{Footnote} The transition temperature, $\mathit{T_{SDW}}$, is suppressed to 182(1) K, 173(2) K, 149(1) K, and 70(2) K, by 1.4\%, 2.7\%, 7.5\%, and 18\% Co substitution, respectively. Previous neutron diffraction measurement revealed that the moment size of Fe is significantly suppressed, from 0.98(8) $\mathit{\mu_{B}}$ in the parent compound to $\sim$0.15(1) $\mathit{\mu_{B}}$ with x = 0.180(5).\cite{Jin_13}  It is worth noting that the temperature dependencies of the (1, 0, 3) peak show a kink around $\mathit{T_{Eu}}$ for the very underdoped compositions with $\mathit{x}$ = 0.014(2) and 0.027(2), suggesting a possible interplay between the Fe and Eu sublattices. In addition, for $\mathit{x}$ = 0.180(5), the behavior that the (1, 2, 1) reflection shows a maximal intensity around its superconducting transition temperature, $\mathit{T_{SC}}$, indicates the competition between SC and the SDW order of Fe, as discussed in details in Ref. \onlinecite{Jin_13}.

\begin{figure}
\centering{}\includegraphics{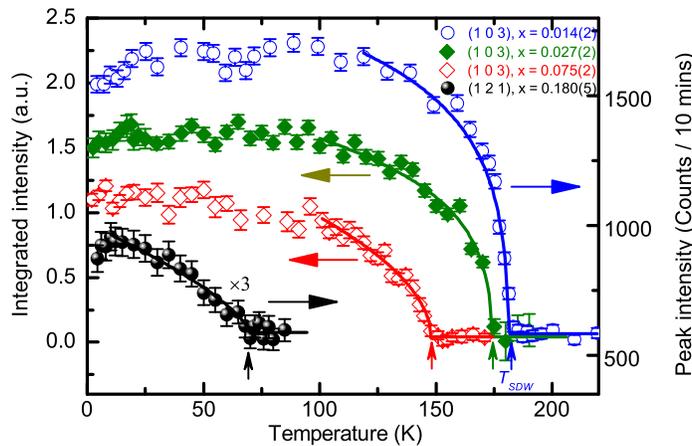}

\caption{The temperature dependencies of the order parameter associated with the SDW transition of Fe in Eu(Fe$_{1-x}$Co$_{x}$)$_{2}$As$_{2}$ single crystals with $\mathit{x}$ = 0.014(2), 0.027(2), 0.075(2) and 0.180(5)\cite{Jin_13}, respectively. The peak intensity of the (1, 0, 3) or (1, 2, 1) magnetic reflection, measured on DNS or TriCS, is plotted for $\mathit{x}$ = 0.014(2) and 0.180(5), respectively. For $\mathit{x}$ = 0.027(2) and 0.075(2), the integrated intensity of the (1, 0, 3) reflection measured on HEiDi or D23 is shown instead.\cite{Footnote} The solid lines are guides to the eye.}
\end{figure}

\section{Discussion And Conclusion}

Combining the results of both macroscopic (the resistivity, magnetic susceptibility and specific heat) and microscopic (neutron diffraction) measurements, we have established the phase diagram of Sn-flux-grown Eu(Fe$_{1-x}$Co$_{x}$)$_{2}$As$_{2}$ single crystals. As shown in Fig. 9, both the structural phase transition ($\mathit{T_{S}}$, determined from resistivity measurements) and the SDW transition of Fe ($\mathit{T_{SDW}}$, determined from neutron diffraction) get continuously suppressed by Co doping. Nevertheless, both of them are still present up to the Co concentration of 18\%. The two transitions become split with increasing doping level, similar to that in Ba(Fe$_{1-x}$Co$_{x}$)$_{2}$As$_{2}$.\cite{Pratt_09, Nandi_10} On the other hand,  the magnetic ground state of the Eu$^{2+}$ spins shows a systematic evolution with Co doping. Consistent with previous studies on Eu(Fe$_{1-x}$Co$_{x}$)$_{2}$As$_{2}$ using  M\"ossbauer spectroscopy and magnetic torque,\cite{Blachowski_11, Guguchia_11_torque}  the magnetic ground state of the Eu$^{2+}$ moments is found to depend strongly on the Co doping level. For relatively low Co doping levels (the area in red), the ground-state magnetic structure of Eu is the A-type AFM structure (Fig. 4(a)) without any net FM moment component, similar to that of the parent compound. For high Co doping levels (the area in blue), the Eu$^{2+}$ moments align in the pure FM order along the crystallographic $\mathit{c}$ axis (Fig. 4(d)). In the intermediate doping levels (the area in gradient colors), the Eu$^{2+}$ spins rotate gradually off the $\mathit{ab}$ plane and form the A-type canted AFM structure, yielding a net FM moment component along the $\mathit{c}$ direction (Figs. 4(b) and 4(c)). The ordering temperature of the Eu sublattice, $\mathit{T_{Eu}}$, declines linearly at first, and then reverses upwards after reaching a minimum value of 16.5(2) K around $\mathit{x}$ = 0.100(4). Note that $\mathit{T_{Eu}}$ reaches 39 K for the end member EuCo$_{2}$As$_{2}$ ($\mathit{x}$ = 1), as reported by Ballinger $et\, al.$\cite{Ballinger_12} The tendency how $\mathit{T_{Eu}}$ develops with Co doping resembles that in EuFe$_{2}$(As$_{1-x}$P$_{x}$)$_{2}$ system,\cite{Cao_11,Jeevan_11,Zapf_11} where a ``dip'' or a minimum was also observed in the $\mathit{T_{Eu}}-x$ curve. The change of the magnetic ground state and the ordering temperature of the Eu sublattice probably arises from the combined effects of the doping-induced modification of the indirect Ruderman-Kittel-Kasuya-Yosida (RKKY) interaction between the Eu$^{2+}$ moments, which is mediated by the conduction $\mathit{d}$ electrons on the (Fe,Co)As layers,\cite{Ruderman_54,Kasuya_56,Yosida_57,Akbari_13} as well as the change of the strength of the direct interaction between the Eu$^{2+}$  and Fe$^{2+}$ moments.  As the RKKY interaction depends on the interlayer distance of the Eu$^{2+}$ moments as well as the details of the Fermi surface, it is therefore expected that by introduction of electron charge carriers and the contraction of the $\mathit{c}$ lattice constant due to the substitution of Co for Fe,\cite{Leithe-Jasper_08} both the strength and the sign of the RKKY exchange coupling can be significantly modified. In addition, the effect due to the change of the Eu-Fe coupling strength tuned by Co doping can not be neglected, since a lot of experimental evidences have revealed a considerable interplay between the localized Eu$^{2+}$ spins and the conduction electrons on the FeAs layers in doped EuFe$_{2}$As$_{2}$.\cite{Ying_10,Guguchia_11,Goltz_14,Jin_16}
 .

\begin{figure}
\centering{}\includegraphics{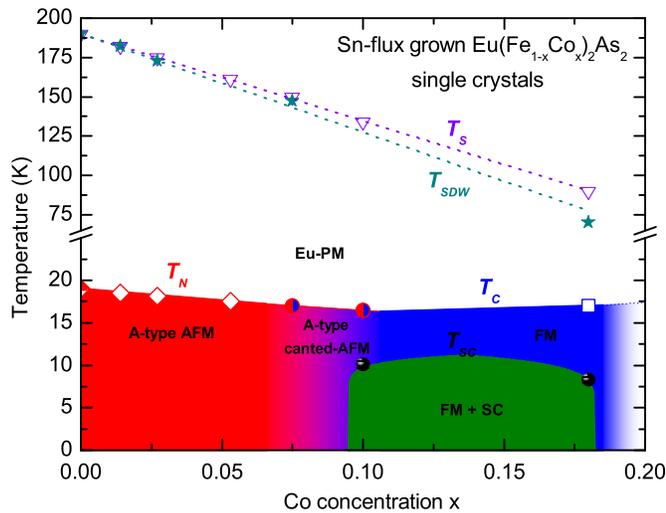}

\caption{The phase diagram of Sn-flux-grown Eu(Fe$_{1-x}$Co$_{x}$)$_{2}$As$_{2}$ single crystals, illustrating the evolution of the magnetic ground state of the Eu$^{2+}$ spins with Co doping. $\mathit{T_{C}}(x)$ is assumed to vary linearly as $\mathit{T_{N}}(x)$ does.  The $\mathit{T_{N}}$ value for the parent compound ($\mathit{x}$ = 0) is obtained from the result of previous neutron diffraction measurement on Sn-flux-grown EuFe$_{2}$As$_{2}$ single crystal (Ref. \onlinecite{Xiao_09}). $\mathit{T_{S}}$   and $\mathit{T_{SDW}}$ denote the structural phase transition and the SDW transtion of Fe, respectively. The dotted lines are linear fittings to $\mathit{T_{S}}$ and $\mathit{T_{SDW}}$.  The right end part is faded because no neutron data are available for that regime.}
\end{figure}

In addition, in the phase diagram of Sn-flux-grown Eu(Fe$_{1-x}$Co$_{x}$)$_{2}$As$_{2}$ single crystals, there is a regime (0.10 $\leqslant$ $\mathit{x}$ $\leqslant$ 0.18, in green color) where strong ferromagnetism from the Eu sublattice is well developed in the superconducting state. Such coexistence of strong ferromagentism and superconducivity was also observed in other doped EuFe$_{2}$As$_{2}$ compounds including EuFe$_{2}$(As$_{1-x}$P$_{x}$)$_{2}$,\cite{Ren_09,Nandi_14,Nandi_14_neutron}
Eu(Fe$_{1-x}$Ru$_{x}$)$_{2}$As$_{2}$,\cite{Jiao_11} and Eu(Fe$_{1-x}$Ir$_{x}$)$_{2}$As$_{2}$.\cite{Jiao_13,Jin_15,Anand_15} Interestingly, in the phase diagram shown in Fig. 9, the onset of superconductivity seems to accompany the formation of strong ferromagnetism. It is intriguing how the two competing phenomena reach a compromise in these compounds. As one possible solution of this puzzle, the existence of a spontaneous vortex state was suggested.\cite{Jiao_11} However, direct evidences for such a state are still lacking and additional measurements such as small angle neutron scattering (SANS) are planned.

In summary, the magnetic ground state of the Eu$^{2+}$ moments in a series of Eu(Fe$_{1-x}$Co$_{x}$)$_{2}$As$_{2}$ single crystals grown from the Sn flux has been investigated by neutron diffraction measurements. Combined with the results from the macroscopic properties (resistivity, magnetic susceptibility and specific heat) measurements, a phase diagram describing how the Eu$^{2+}$ magnetic order evolves with Co doping is established. The ground-state magnetic structure of the Eu$^{2+}$ spins is found to develop from the A-type antiferromagnetic (AFM) order in the parent compound, via the A-type canted AFM structure with some net ferromagnetic (FM) moment component along the crystallographic $\mathit{c}$ direction at intermediate Co doping levels, finally to the pure FM order at relatively high doping levels. The ordering temperature of Eu declines linearly at first, reaches the minimum value of 16.5(2) K around $\mathit{x}$ = 0.100(4), and then reverses upwards with further Co doping. The doping-induced modification of the indirect RKKY interaction between the Eu$^{2+}$ moments, which is mediated by the conduction $\mathit{d}$ electrons on the (Fe,Co)As layers, as well as the change of the strength of the direct interaction between the Eu$^{2+}$ and Fe$^{2+}$ moments, might be responsible for the change of the magnetic ground state and the ordering temperature of the Eu sublattice. . In addition, for Eu(Fe$_{1-x}$Co$_{x}$)$_{2}$As$_{2}$ single crystals with 0.10 $\leqslant$ $\mathit{x}$ $\leqslant$ 0.18, strong ferromagnetism from the Eu sublattice is well developed in the superconducting state, where a spontaneous vortex state is proposed to account for the compromise between the two competing phenomena. 

\bibliographystyle{apsrev} \bibliographystyle{apsrev}

\begin{thebibliography}{10}
\bibitem{Footnote}
Unfortunately, the order parameters for the other two compositions were not able to be measured because of the extreme weakness of the magnetic reflections of Fe, due to the smallness of the crystal with $\mathit{x}$ = 0.053(2) and the significant neutron-absorption effect with cold neutrons for $\mathit{x}$ = 0.100(4), in which the Fe-SDW order is already suppressed to be quite weak.
\end{thebibliography}
\begin{acknowledgments}
This work is partly based on experiments performed at the HEIDI and DNS instrument operated by J\"{u}lich Centre for Neutron Science (JCNS) at the Heinz Maier-Leibnitz Zentrum (MLZ), Garching, Germany. It is also partly based on experiments performed at the Swiss Spallation Neutron Source (SINQ), Paul Scherrer Institute, Villigen, Switzerland, and at the Institut Laue-Langevin (ILL), Grenoble, France. W. T. J. would like to acknowledge S. Mayr and J. Persson for their assistance for the orientation of the crystals. Z.B. acknowledges financial support from the National Science Center of Poland, Grant No. 2011/01/B/ST5/06397. 
\end{acknowledgments}

\end{document}